\documentclass[aps,prb,twocolumn,groupedaddress,superscriptaddress,showpacs]{revtex4}
\pdfoutput=1
\usepackage{amsmath}
\usepackage{graphicx}
\usepackage{bm}

\newcommand{\Hca}{\mathcal{H}}

\newcommand{\la}{\langle}
\newcommand{\ra}{\rangle}

\usepackage{color}
\usepackage{color}

\newcommand{\zb}{\bar{z}}

\DeclareMathOperator{\sign}{sign}
\DeclareMathOperator{\real}{Re}
\DeclareMathOperator{\imag}{Im}

\newcommand{\noo}[1]{: \negthinspace {#1} \negthinspace :}

\begin{document}

\title{Theory of non-Abelian Fabry-Perot interferometry in topological insulators}

\author{Johan Nilsson}
\affiliation{Department of Physics, University of Gothenburg,
412 96 Gothenburg,  Sweden}
\affiliation{Instituut-Lorentz, Universiteit Leiden,
P.O. Box 9506, 2300 RA Leiden, The Netherlands}

\author{A. R. Akhmerov}
\affiliation{Instituut-Lorentz, Universiteit Leiden,
P.O. Box 9506, 2300 RA Leiden, The Netherlands}

\date{December 2009}

\begin{abstract}
Interferometry of non-Abelian edge excitations is a useful tool in topological quantum computing. In this paper we present a theory of a non-Abelian edge state interferometry in a 3D topological insulator brought in proximity to an s-wave superconductor. The non-Abelian edge excitations in this system have the same statistics as in the previously studied 5/2 fractional quantum Hall (FQH) effect and chiral p-wave superconductors. There are however crucial differences between the setup we consider and these systems, like the need for a converter between charged and neutral excitations and the neutrality of the non-Abelian excitations. These differences manifest themselves in a temperature scaling exponent of $-7/4$ for the conductance instead of $-3/2$ as in the 5/2 FQH effect.
\end{abstract}


\maketitle

\section{Introduction}

One of the most promising tools in topological quantum computing\cite{Kitaev_top_computing_2003,Nayak_RMP_topological_2008} is non-Abelian edge state interferometry.\cite{DasSarma_topcharge2005,SternHalperin2006,Bonderson2006} Its main idea is that moving a fractional excitation (anyon) existing at an edge of a topological medium around localized anyons in the bulk allows to extract information about the state of the latter. The theory of edge state interferometry was initially developed for Ising anyons in the 5/2 fractional quantum Hall (FQH) state and p-wave superconductors,\cite{DasSarma_topcharge2005,SternHalperin2006,Bonderson2006,Fendley2007a} building on earlier work on FQH systems.\cite{Chamon1997a,Fradkin_1998} Recent experiments,\cite{Willet_PNAS_2009} which provide evidence for non-Abelian braiding statistics in the 5/2 FQH state (see the detailed discussion in Ref.~\onlinecite{Bishara_interferometer_2009}) are using this method, and it is generally considered the most promising way to measure the state of topological qubits.

\begin{figure}[htb]
\centering
\includegraphics[scale=.5]{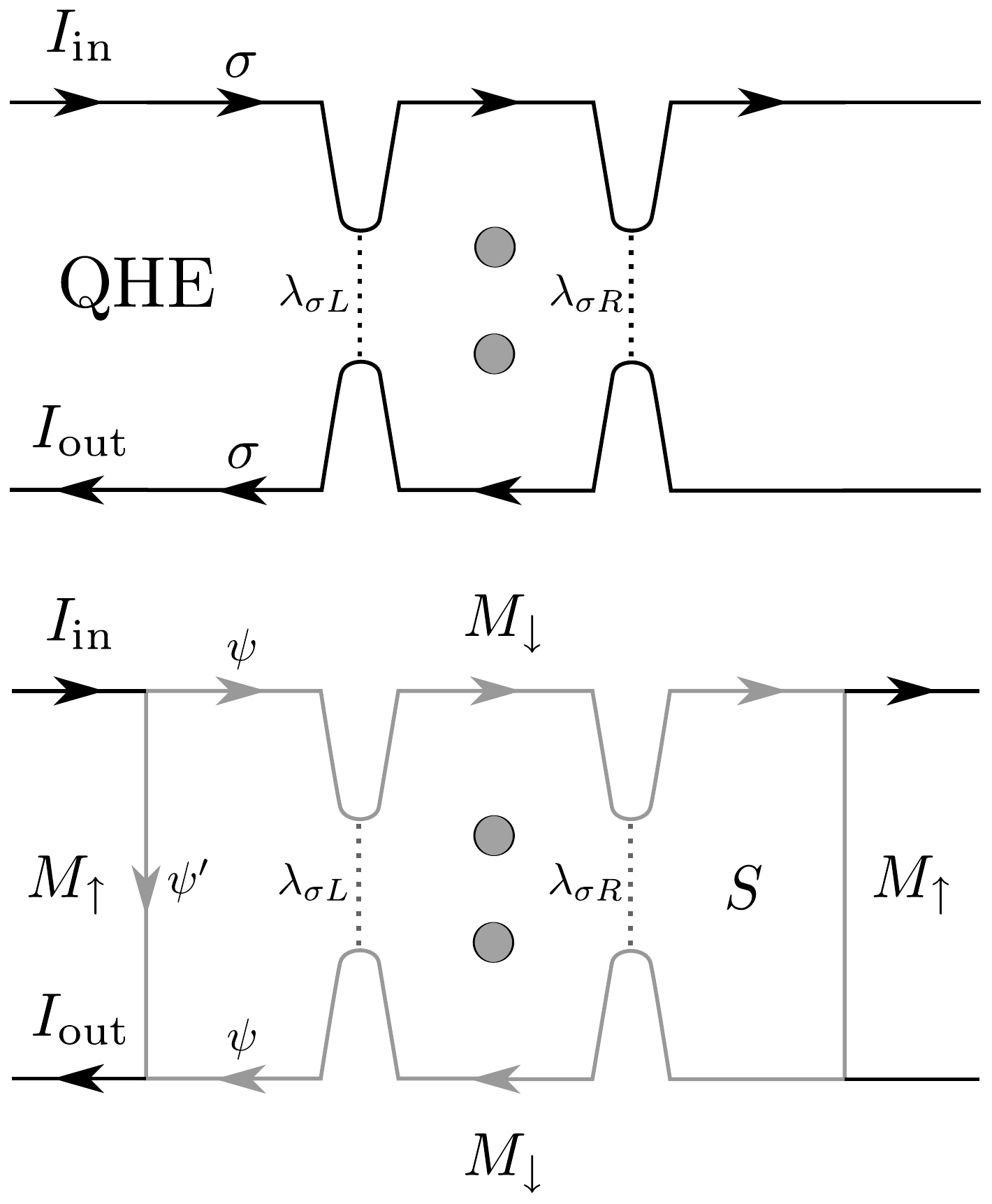}
\caption{\label{fig:setup} Edge state Fabry-Perot interferometer in the 5/2 FQH system (top panel) and in a topological insulator/s-wave superconductor heterostructure (bottom panel). The charge is transferred locally at the tunneling point in FQH effect, and is only well-defined in the ferromagnetic domain walls (i.e. the leads) in the topological insulator setup. Regions labeled $S$, $M_\uparrow$, and $M_\downarrow$ denote parts of topological insulator in proximity of a superconductor and of ferromagnetic insulators with different polarizations. Grey circles in the middle of the central island are Majorana bound states forming a Majorana qubit, which can be measured by the interferometer.}
\end{figure}

We present a theory of non-Abelian edge state interferometry of the Majorana modes existing at the surface of a 3D topological insulator brought in contact with an s-wave superconductor and a ferromagnetic insulator.\cite{FuKane_PRL2008} The main difference of an interferometry setup in this system, as compared with 5/2 FQH interferometer, is the need for an additional ``Dirac to Majorana converter''.\cite{FuKane_PRL2009,Akhmerov_PRL_2009} This element is required because unlike in the FQH effect the edge excitations near a superconductor carry no charge and thus allow no electric readout. This converter initially transforms the charged excitations injected from a current source into superpositions of two neutral excitations existing at different edges of the superconductor.  Later another converter recombines a pair of neutral excitations exiting the interferometer into a charged particle, either an electron or a hole, that can be measured as a current pulse. The difference between the two systems is summarized in Fig.~\ref{fig:setup}. The ``Dirac to Majorana converter'' is not available in chiral p-wave superconductors, since the chirality of the neutral edge modes is then set by time-reversal symmetry breaking in the condensate, and not by the external region of the system (magnet). Such a limitation combined with the absence of charged modes makes electric readout of interferometry experiment much less viable in a chiral p-wave superconductor.

The description of the ``Dirac to Majorana converter'' using single particle formalism was done in Refs.~\onlinecite{FuKane_PRL2009,Akhmerov_PRL_2009}. The qualitative description of the non-Abelian Fabry-Perot interferometer was presented in Ref.~\onlinecite{Akhmerov_PRL_2009}. In this paper we use conformal field theory (CFT) to describe and analyze the non-Abelian excitations following Ref.~\onlinecite{Fendley2007a}. 

An important difference between the systems is the following: In the 5/2 FQH effect the charge density and accordingly charge current of anyons may be defined locally, since anyons have charge $e/4$ or $e/2$ in this system. Excitation of charge $e^*$ has an energy cost of $e^* V$ for being created in the system. This energy cost provides a natural cutoff for the current, whereas in the superconducting systems due to the absence of charge in the edge excitations the only cutoff is set by the finite temperature. The neutrality of the edge excitations does not only mean that a finite voltage does not provide a cutoff for the conductance, but also results in a different temperature scaling exponent of the conductance. In the topological insulator setup the conductance diverges at low temperatures as $\Gamma\sim T^{-7/4}$, while in the FQH setup it goes as $\Gamma \sim T^{-3/2}$.

The experimental requirements for a realization of edge state interferometry in topological insulators were discussed in Refs.~\onlinecite{FuKane_PRL2009} and \onlinecite{Akhmerov_PRL_2009}. An additional requirement for non-Abelian interferometry is the need for a sufficiently high amplitude of the vortex tunneling, $\lambda_\sigma \sim \exp(-\sqrt{E_C/E_J})$, with $E_J$ the Josephson energy and $E_C$ the charging energy. It is non-negligible only if the superconducting islands in the system have small capacitive energy $E_C$.\cite{Mooij_Nazarov2006}

The outline of this paper is as follows: In Sec.~\ref{sec:chiralfermions} we introduce the effective model that we use to describe the fermions that propagate along magnetic domain walls and the superconducting-magnet domain walls. In particular we introduce the representation of these fermions in terms of Majorana fields, which we use later. In Sec.~\ref{sec:linearresponse} we review the linear response formula that we use to calculate the non-local conductance, the experimentally relevant quantity that we are interested in. In Sec.~\ref{sec:perturbative} we give a detailed account of the perturbative calculation of the conductance, and we consider the most interesting case of vortex tunneling in Sec.~\ref{sec:vortextunneling}. In Sec.~\ref{sec:topologicalcharge} we show how the proposed setup can be used to measure the fermion parity (and hence the topological charge) of the Majorana qubit that is stored in a pair of bulk vortices. Our conclusions are to be found in Sec.~\ref{sec:conclusions}. We provide a detailed description of the formalism that we use to describe the peculiar vortex field in the appendices.

\section{Chiral fermions}
\label{sec:chiralfermions}

\subsection{Domain wall fermions}

It is known that there exists a single chiral fermion mode on each mass domain wall in the 2D Dirac equation. This mode is localized near the domain wall but is allowed to propagate along the domain wall in only one direction (hence the name chiral). This is most easily seen using an index theorem that relates the difference in a topological number ($\tilde{N}_3$ in the language of Ref.~\onlinecite{Volovik_Book}) between the two domains and the difference in the number of right- and left-moving states that live in the domain wall.\cite{Volovik_Book} In the ferromagnetic domain wall that we are interested in the change in $\tilde{N}_3$ across the domain wall is $\pm 1$. If the domain wall is also abrupt enough then only one chiral fermion exists in the domain wall.

A similar argument can be made using the Dirac-Bogoliubov-de Gennes (BdG) equation with gaps generated by the superconducting order parameter  $\Delta$. In the case that we consider (s-wave pairing) $\tilde{N}_3$ is zero if the gap is dominated by the superconducting gap $|\Delta|$ and non-zero ($\pm 1$) when the gap is of ferromagnetic character. Because of the double counting of states in the BdG equation this implies that $\frac{1}{2}$ of a chiral fermion state exists on a superconducting-magnetic domain wall. This is exactly the number of degrees of freedom that is encoded in a {\it chiral Majorana fermion} field.

Alternatively one can argue for the existence of these states by solving the BdG equation explicitly for certain simple domain wall profiles or use $\bm{k \cdot p}$ theory.\cite{FuKane_PRL2009} We now proceed to a theoretical description of these states. In particular we will see that it is fruitful to describe both kinds of domain walls in terms of Majorana fields.

\subsection{Theoretical description}

In the leads (ferromagnetic domain walls), where the superconducting order parameter vanishes, the system consists of a single normal edge state which propagates in only one direction, i.e. a single chiral charged mode. This can be described by a complex fermionic field $\hat{\Psi}(x)$ with Hamiltonian
\begin{equation}
H(t) = \frac{1}{2\pi} \int dx \noo{ \hat{\Psi}^\dagger(x) [ v p_x - \mu(x,t)]  \hat{\Psi}(x) }.
\label{eq:Hnormaledge}
\end{equation}
Here $: \; :$ denotes normal ordering.
We use units such that $\hbar =1$ unless specified otherwise. The kinetic energy operator $v p_x$ is defined as
\begin{equation}
v p_x = i \frac{\overset{\leftarrow}{\partial_x}v(x) - v(x) \overset{\rightarrow}{\partial_x}}{2}
\rightarrow -i \sqrt{v(x)} \overset{\rightarrow}{\partial_x} \sqrt{v(x)},
\end{equation}
where we have introduced the spatially varying velocity $v(x)$ in a symmetric way such that $v p_x$ is a Hermitean operator. The stationary (energy $E$) solution to the time-dependent Schr\"odinger equation corresponding to Eq.~\eqref{eq:Hnormaledge} for zero chemical potential $\mu = 0$ is
\begin{equation}
\Psi_E(x,t) = \sqrt{\frac{v(0)}{v(x)}} \exp \Bigl(i E \Bigl[ \int_0^x \frac{dx'}{v(x')}  -t  \Bigr] \Bigr) \Psi_E(0,0).
\end{equation}
This implies that
\begin{equation}
\la \hat{\Psi}(x,t) \hat{\Psi}^\dagger(0,0) \ra = 
\frac{[v(x) v(0)]^{-1/2}}{a + i \bigl[ t- \int_0^x dx' / v(x') \bigr]},
\label{eq:varyingvcorrelation}
\end{equation}
where $a$ is a short time cutoff which should be taken to zero. If the velocity $v$ is constant the result simplifies to
\begin{equation}
v \la \hat{\Psi}(x,t) \hat{\Psi}^\dagger(0,0) \ra =  \frac{1}{a + i (t - x/v)} \equiv \frac{1}{a + i u}.
\end{equation}
The normalization in Eq.~\eqref{eq:Hnormaledge} is chosen to yield this result without any extra normalization factors. Note that it implies (in the limit $a\rightarrow 0^+$) that the anti-commutation relation for the field is $\{\hat{\Psi}(x),\hat{\Psi}^\dagger(x')\} = 2 \pi \delta(x-x')$.

An important consequence of the chiral nature of the excitations is that the correlation functions only depend on the difference of the {\it Lorentz time} $u = t -x/v$. According to Eq.~\eqref{eq:varyingvcorrelation} the same is true also for a spatially varying velocity with the proper interpretation of the length difference. Because of this property we will mostly work with a spatially homogeneous velocity that we will set to unity ($v=1$) in the following calculations. It is also useful to go from the Hamiltonian to the corresponding Lagrangian
\begin{equation}
L = \frac{1}{2\pi} \int dx \noo{ \hat{\Psi}^\dagger(x) [ i \partial_ t - v p_x + \mu(x,t)] \hat{\Psi}(x) },
\label{eq:Snormaledge}
\end{equation}
since the coupling to the gauge field is most transparent in this formalism.

\subsection{Majorana fermion representation}

We can decompose $\hat{\Psi} (x)$ into two independent Majorana fields $\psi(x) = \psi^\dagger(x)$ and $\psi'^\dagger(x) = \psi'(x)$ as
\begin{equation}
\hat{\Psi}(x,t)  = \frac{e^{i A(x,t)}}{\sqrt{2}} [\psi (x,t) + i \psi' (x,t)].
\end{equation}
The anti-commutation relations of the Majorana fields are $\{ \psi(x), \psi(x') \} = \{ \psi'(x), \psi'(x') \} =2 \pi \delta(x-x')$, and $\{ \psi(x), \psi'(x') \} = 0$. In terms of $\psi$ and $\psi'$ the Lagrangian becomes
\begin{multline}
L =  \frac{1}{4\pi}\int dx 
\bigl[ \noo {\psi (x) (i\partial_t - v p_x )\psi (x) }
\\ +
 \noo {\psi' (x) (i\partial_t - v p_x ) \psi' (x) } \bigr]
\\ +
\frac{i e}{2\pi} \int dx F(x,t)  v(x) \psi' (x) \psi (x),
\label{eq:majoranahamiltonian0}
\end{multline}
where $F(x,t)$ depends on the phase $A(x,t)$, i.e. it is gauge dependent:
\begin{equation}
-e F(x,t) = \frac{\mu(x,t)}{v(x)} - \frac{1}{v(x)}\partial_t A(x,t) - \partial_x A(x,t) .
\end{equation}
Note that this means that a time-independent spatially varying chemical potential can be gauged away up to possible boundary terms.

\begin{figure}[htb]
\centering
\includegraphics[scale=.5]{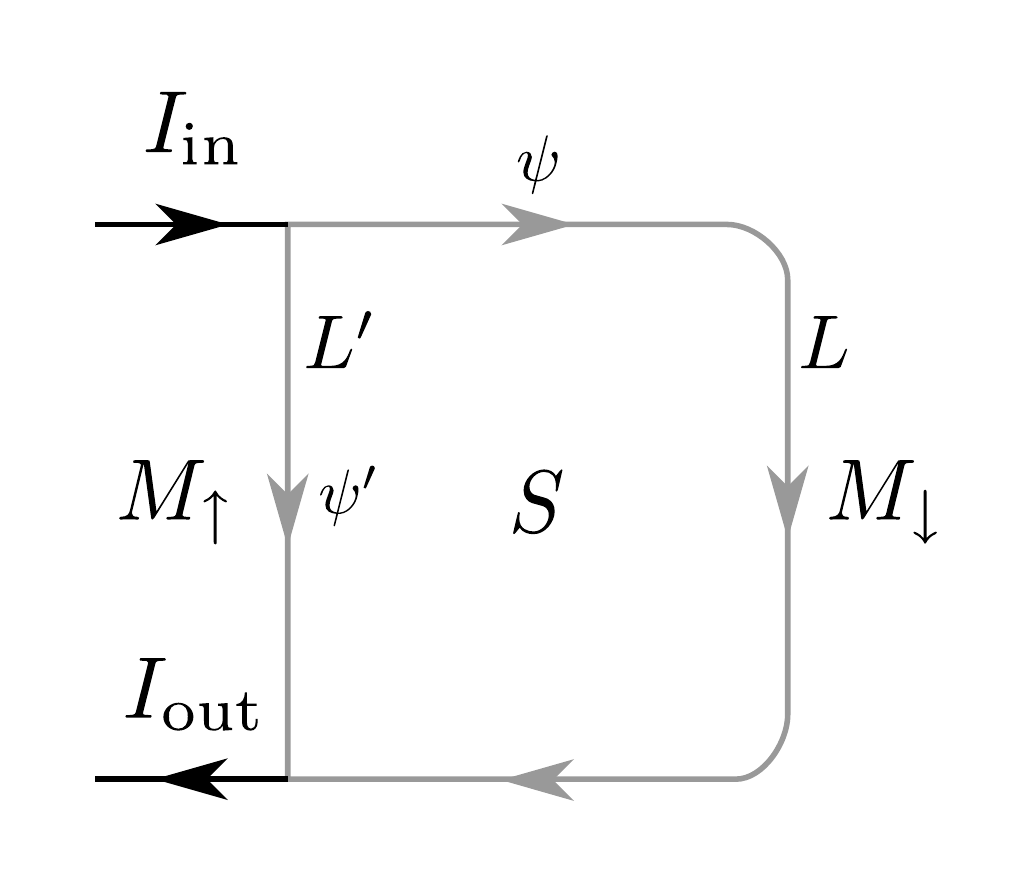}
\caption{Free fermion propagation setup. The two Majorana modes $\psi$ and $\psi'$ are spatially separated by the superconducting region. Thus the effective propagation length from in to out can be different for the two modes, i.e. $L' \neq L$.}
\label{fig:setup_free_propagation}
\end{figure}

One of the most interesting features of the system that we consider is that the two Majorana fields that appear in this action can becomes spatially separated when a superconducting region is sandwiched in between the two magnetic regions in a magnetic domain wall as discussed previously. Thus the action in Eq.~\eqref{eq:majoranahamiltonian0} can be used to describe the setup in Fig.~\ref{fig:setup_free_propagation}, in which the two Majorana fields $\psi$ and $\psi'$ are spatially separated inside of the interferometer. It is important to remember that the coordinate systems of the two fields are different in this representation.

From the Lagrangian and the coupling to the gauge field  we now identify the charge current operator as
\begin{equation}
\hat{J}(x) = \frac{-e v(x)}{2\pi} \noo{ \hat{\Psi}^\dagger(x) \hat{\Psi}(x) }  = \frac{i e}{2\pi} v(x) \psi' (x) \psi(x).
\label{eq:currentop}
\end{equation}
This form of the current operator in terms of the Majorana fields is very important for the following calculations. It is only well-defined if the two Majorana modes are at the same position in space, hence there is no coupling to the electric field inside of the interferometer where the two Majorana wires are spatially separated. This is also an important difference between the FQH setup where local charge current operators can be defined at the tunneling point contacts. This simplifies the calculation because the local charge transfer is directly related to the measurements done far away. In our system we don't have this luxury and must consider the leads explicitly.

\section{Linear response formalism for the conductance}
\label{sec:linearresponse}

If we write the Lagrangian in Eq.~\eqref{eq:majoranahamiltonian0} as $L = L_0 - H'(t)$, where the term on the last line is
\begin{equation}
H'(t) = -  \int dx \hat{J} (x,t) F(x,t),
\end{equation}
we are in the position to use the standard linear response Kubo formula,\cite{fetterwalecka} to calculate the conductance tensor $\Gamma$. Following Ref.~\onlinecite{BarangerStone1989} we introduce an AC chemical potential localized in the source lead, which we take to have coordinates $x<0$. We choose a constant gauge $A(x,t) = A$ so that $F(x,t)= -\Theta(-x) \cos (\Omega t) e^{- \delta |t|}V / v(x)$.\cite{gauge_choice_comment1} The conductance $\Gamma$ is defined as the magnitude of the in-phase current divided by the applied voltage difference $V$. Following the usual steps, with the current operator in Eq.~\eqref{eq:currentop} and assuming that the two Majorana modes are independent, we obtain the formula
\begin{multline}
\Gamma = \frac{e^2}{\pi h} \lim_{\Omega,\delta \rightarrow 0^+}
\int_0^{\infty} dt'  \int_0^\infty dt  
\imag [ G^>_{ji} {G}^>_{j'i'} ] \cos (\Omega t) e^{- \delta t}.
\label{eq:Gamma_v1}
\end{multline}
Here we have reintroduced the correct units of conductance $e^2/h$. We have also used the fact that in a chiral system the response in the region $x>0$ to a spatially uniform extended source $x \leq 0$ at a particular time $t'=0$ is equivalent to the response to a point source at $x=0$ that is on for $t' \geq0$. The important quantities to calculate are the Green's functions
\begin{subequations}
\begin{eqnarray}
G^>_{ji}  &\equiv& \la \psi (y,t) \psi (0,t') \ra \equiv \la \psi_j \psi_i \ra,
\\
G^>_{j'i'}  &\equiv& \la  \psi' (y',t) \psi' (0',t') \ra \equiv \la \psi'_{j' }\psi'_{i'} \ra.
\end{eqnarray}
\end{subequations}
Here the indexes $i$ and $j$ are shorthands for the coordinates of the source $(0,t')$ and current measurement $(y,t)$. Similarly for the primed coordinate system, which is typically not the same in the setups that we consider as discussed previously.

Because the correlation functions only depends on $t-t'$ it is possible to perform the integral over $t+t'$ in Eq.~\eqref{eq:Gamma_v1} explicitly, the resulting expression is
\begin{equation}
\Gamma = 
- \frac{e^2}{\pi h}  \int_0^\infty dt 
\imag [ G^>_{ji} {G}^>_{j'i'} ] t ,
\label{eq:Gamma_v2}
\end{equation}
where it is understood that the source term is taken at $t'=0$.
Here we have also used the fact that the correct limit is to take $\delta \rightarrow 0^+$ first and then $\Omega \rightarrow 0$. Because we are interested in the finite temperature result the cut-off provided by the thermal length is enough to render the expression convergent. This is the master formula that we will use to calculate the conductance in the following.

\subsection{Free fermion propagation}
\label{sec:freefermionprop}

If both Majorana modes propagates freely (the setup is sketched in Fig.~\ref{fig:setup_free_propagation}) we can use the finite temperature propagator
\begin{multline}
G^>_{j'i'} =\frac{1}{z_{j'i'}} \equiv \frac{\pi T}{\sin \pi T [a + i u_{j'i'}]  }
\\
\underset{a \rightarrow 0^+}{=}
 \pi \delta (u_{j'i'}) - i \mathcal{P} \frac{\pi T}{\sinh(\pi T u_{j'i'})},
\label{eq:Gprime0}
\end{multline}
where $u_{j'i'} = t - L'$. The Green's function of the other edge $G^>_{ji}$ is given by the same expression with $L$ (the effective length of propagation) instead of $L'$. Substituting the expressions for the Green's functions into Eq.~\eqref{eq:Gamma_v2} we obtain
\begin{equation}
\Gamma = \frac{e^2}{h} \frac{\pi T (L-L')}{\sinh[\pi T (L-L')]},
\end{equation}
in the limit $a \rightarrow 0^+$. This formula agrees with the linear response limit of the the result obtained with the scattering formalism in Ref.~\onlinecite{FuKane_PRL2009}, and shows how the path difference enters in the finite temperature case.

To obtain the response in the source lead we take the limit $L' \rightarrow L$ with the result that
\begin{equation}
\Gamma = \frac{e^2}{h} .
\label{eq:onechannel}
\end{equation}
This is the expected (and correct) result for a system with one propagating channel. If $L' \neq L$ we also obtain Eq.~\eqref{eq:onechannel} as long as $T |L-L'| \ll 1$, in the zero temperature limit the result is thus independent of the path length difference. The Eq.~\eqref{eq:onechannel} agrees with the limit $V \rightarrow 0$ of the previous results,\cite{FuKane_PRL2009,Akhmerov_PRL_2009} which were based on the scattering formalism.

\section{Perturbative formulation}
\label{sec:perturbative}

In tunneling problems we want to calculate the Green's function
$G^>_{ji} = \la \psi_j \psi_i \ra$, where $\psi_i$ and $\psi_j$ live on different edges of the sample, in the presence of a perturbation $\delta H$ that couples the two edges. Assuming that the system is in a known state at time $t_0$, we may express the expectation value in the interaction picture as
\begin{equation}
G^>_{ji} = \la U(t_0,t) \psi_j(t) U(t,0) \psi_i (0) U(0,t_0)\ra.
\end{equation}
Here $U(t,t')$ is the time evolution operator in the interaction picture. For $t \geq t'$ it is given by the familiar time-ordered exponential $U(t,t') = \mathcal{T} \exp[-i \int_{t'}^t ds \delta H(s)]$.

In the following we will assume that the average at $t=t_0$ is a thermal one at temperature $T$. A perturbative expansion is obtained by expanding the time-ordered and anti-time-ordered exponentials in this expression in powers of $\delta H$. This procedure is equivalent to the Schwinger-Keldysh formalism, which in addition provides a scheme to keep track of whether one is propagating forward or backward in time. We will also assume that the perturbation was turned on in the infinite past, i.e. we set $t_0 = -\infty$. 

\subsection{Fermion tunneling}
\label{subs:fermion_tunneling}
\begin{figure}[htb]
\centering
\includegraphics[scale=.5]{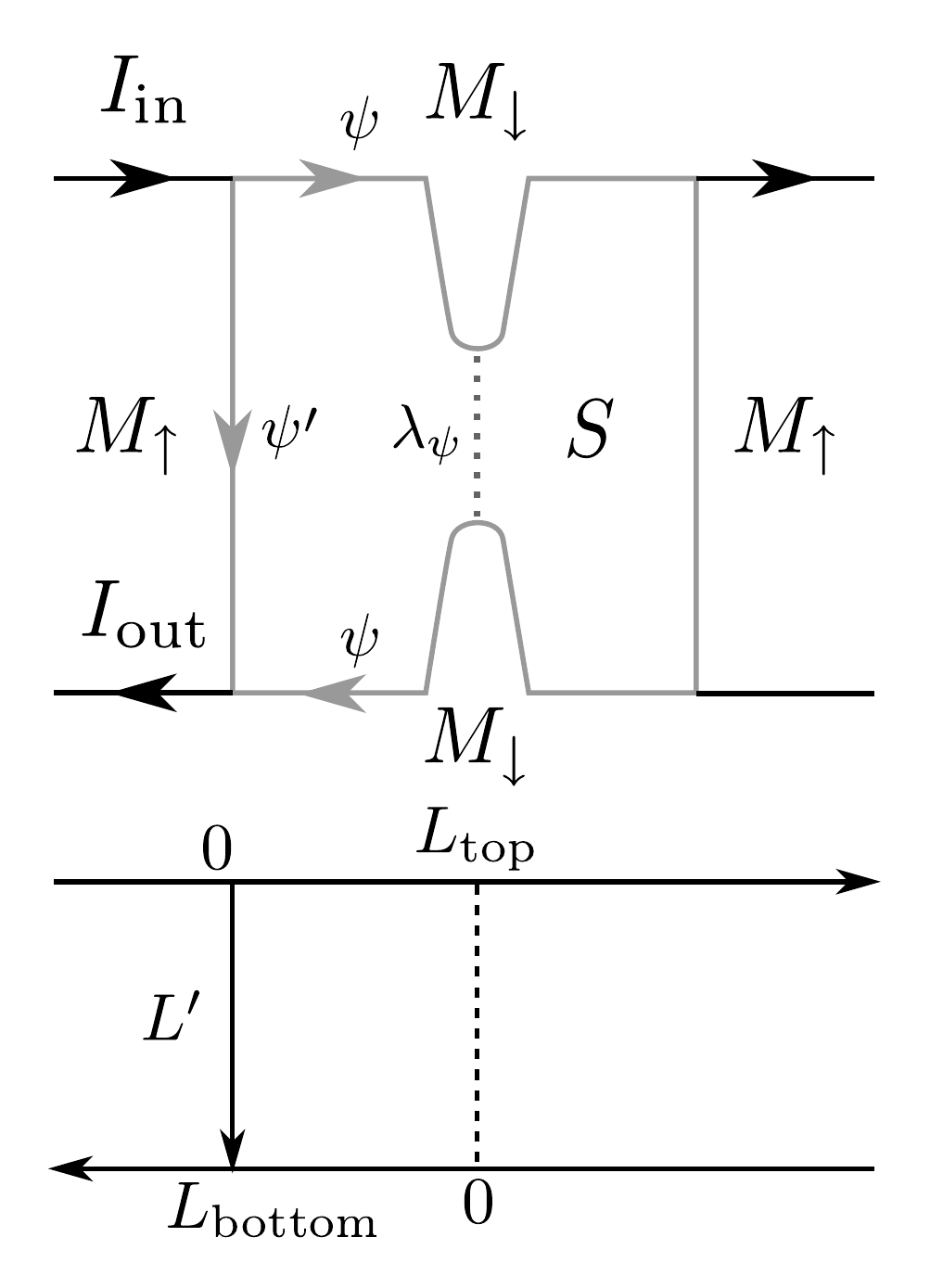}
\caption{Top panel: fermion tunneling setup. The coordinate conventions used in Sec.~\ref{subs:fermion_tunneling} are shown in the bottom panel.}
\label{fig:setup_fermiontunnel}
\end{figure}

As a warm-up for the vortex tunneling calculation we will now consider the simpler case of fermion tunneling, which we describe by a tunneling term $H_\psi(t_1) = i \lambda_\psi \psi_2 \psi_1 / (2\pi)$.\cite{Fendley2007a} Here $\psi_1$ ($\psi_2$) is located at the tunneling point at the upper (lower) edge. The system and the coordinate convention we use are sketched in Fig.~\ref{fig:setup_fermiontunnel}. The leading contribution to conductance comes at first order in the tunneling amplitude $\lambda_\psi$. After a straightforward expansion and collection of terms we obtain
\begin{multline}
G^>_{ji} = \frac{\lambda_\psi}{2\pi} \int_{-\infty}^t dt_1 \{\psi_j , \psi_2 \} \la \psi_1  \psi_i \ra
\\
- \frac{\lambda_\psi}{2\pi} \int_{-\infty}^0 dt_1 \{\psi_i , \psi_1 \} \la \psi_j  \psi_2 \ra
+ \mathcal{O}(\lambda_\psi^2).
\label{eq:fermion_tunnel_calc}
\end{multline}
Here we have used the fact that the two groups of fermions on different edges, i.e ($\psi_j$,$\psi_2$) and ($\psi_i$,$\psi_1$), are independent. It is straightforward  to evaluate this expression using Eq.~\eqref{eq:Gprime0} together with $\{ \psi_i, \psi_1 \} =2 \pi \delta(u_{1i})$, and $\{ \psi_j, \psi_2 \} =2 \pi \delta(u_{2j})$, where $u_{1i} = t_1 - L_{\textrm{top}}$ and $u_{2j} = t_1- t + L_{\textrm{bottom}}$. Because of the geometry of the problem the second term on the right hand side of Eq.~\eqref{eq:fermion_tunnel_calc} vanishes due to causality (the Lorentz time arguments never coincide). The Green's function $G^>_{ji}$ to leading order in tunneling strength is therefore 
\begin{equation}
G^>_{ji} = \lambda_\psi \frac{\pi T}{\sin \pi T [a + i (t-L)]  },
\end{equation}
where $L = L_{\textrm{top}} +L_{\textrm{bottom}}$ is the effective propagation length of the Majorana fermion. Using the result of Sec.~\ref{sec:freefermionprop} we then find that the conductance of this setup is
\begin{equation}
\Gamma =\lambda_\psi \frac{e^2}{h},
\end{equation}
at $T=0$. Once again this result agrees with the zero frequency, zero voltage limit of the results obtained with the scattering method in previous work.\cite{FuKane_PRL2009,Akhmerov_PRL_2009}

\section{Vortex tunneling}
\label{sec:vortextunneling}

The main focus of this paper is to study how the tunneling of a pair of vortices can effectively transfer a fermion, and hence give a contribution to the conductance. Schematically the vortex tunneling term can be written as
\begin{equation}
\Hca_{\sigma} = \lambda_\sigma \sigma_b(x) \sigma_t(x'),
\end{equation}
where the index $t$ ($b$) denotes the top (bottom) edge. As it stands this term is not well-defined without more information about the two spin fields $\sigma$, this is discussed in great detail in Ref.~\onlinecite{Fendley2007a}. We provide a detailed description of the formalism that we use to deal with this issue in the appendixes.

\subsection{Coordinate conventions}
\label{sec:coordinates}

\begin{figure}[htb]
\centering
\includegraphics[scale=.4]{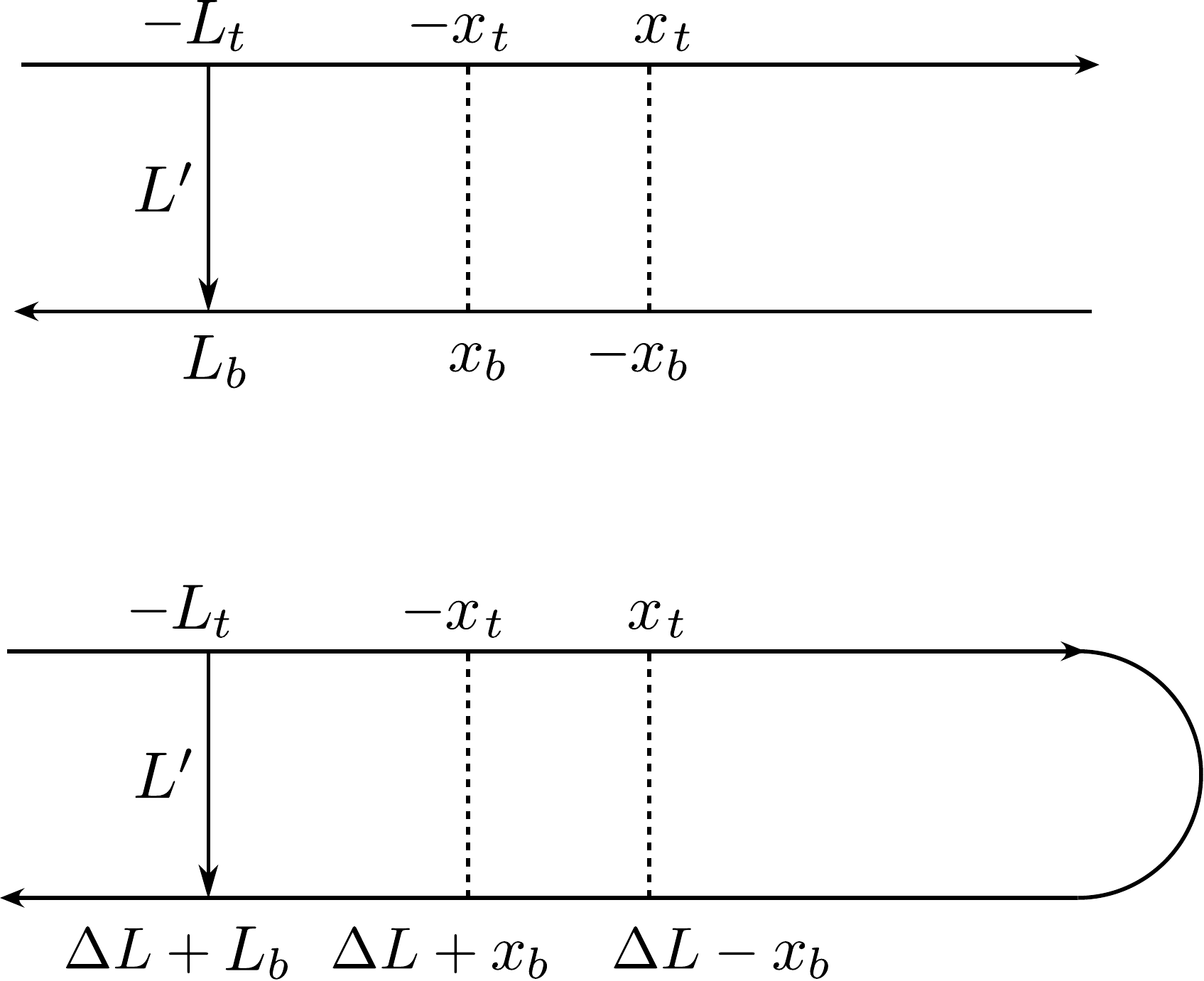}
\caption{Top panel: independent coordinate system for the two edges. Bottom panel: coordinate system in which
the two edges are treated as spatially separated parts of the same edge. This allows us to correctly capture the commutation relations of the fields on different edges in the relevant limit $\Delta L \rightarrow \infty$.}
\label{fig:singe_edge}
\end{figure}

To have a well-defined prescription for the commutation relation of fields on different edges we will treat the two edges as spatially separated parts of the same edge. This reasoning has been employed in a number of works studying tunneling in the FQH effect, see for example Refs.~\onlinecite{Guyon2002a} and \onlinecite{Kim2006a}. This approach leaves a gauge ambiguity: should we choose the bottom edge to have spatial coordinates smaller or larger than that of the top edge? The correct choice is fixed by noting that the current operator at the source should commute with the vortex tunneling term at equal times because of the locality and gauge invariance. A similar argument can be made considering the current operator at the measurement position before the information about the tunneling event has had time to reach it. Since we want the vortex tunneling event to commute with fermions on the reference edge at all times we are forced to use the coordinate convention shown in Fig.~\ref{fig:singe_edge} in which the spatial coordinates on bottom edge are always larger than those on top edge.\cite{gauge_choice_comment2} The vortex tunneling then corresponds to changing the phase of the superconducting order parameter by $\pm 2\pi$ to the right of the tunneling point in the figure.

In addition it is convenient to introduce an even more compact notation. We denote $\psi_t(0,-L_t) \equiv \psi_i$, $\psi_b(t,\Delta L + L_b) \equiv \psi_j$, $\sigma_t(t_1,-x_t) \equiv \sigma_1$, $\sigma_b(t_1,\Delta L +x_b) \equiv \sigma_2$, $\sigma_t(t_2,x_t) \equiv \sigma_3$, $\sigma_b(t_2,\Delta L -x_b) \equiv \sigma_4$. The two tunneling terms in the Hamiltonian are then written as $\lambda_\sigma T_{12}$ and $\lambda_\sigma T_{34}$. The modification needed to allow for different tunneling amplitudes $\lambda_{\sigma L}$ and $\lambda_{\sigma R}$ at the left and right tunneling points (see Fig.~\ref{fig:setup}) is straightforward. The ``Lorentz times'' $u$ for right-movers are $u \equiv t -x$. We use additional short-hand notations $u_{\alpha\beta} \equiv u_\alpha -u_\beta$ and $s_{\alpha \beta} \equiv \sign(u_\alpha - u_\beta)$. The Lorentz times of the six operators used in the calculation are
\begin{eqnarray}
u_{i} &=&  L_t \nonumber  \\
u_{1} &=& t_1 + x_t \nonumber  \\
u_{3} &=& t_2 - x_t   \\
u_{j} &=& t - L_b -\Delta L \nonumber  \\
u_{2} &=&  t_1 - x_b - \Delta L \nonumber  \\
u_{4} &=&  t_2 + x_b -\Delta L. \nonumber
\end{eqnarray}
Taking the limit of large spatial separation $\Delta L \rightarrow + \infty$ we see that $s_{ij} =1$. Accordingly, in this limit also $s_{kl} = 1$ for any $k \in \{i,1,3\}$ and $l \in \{j,2,4\}$.

In the following perturbative treatment we will assume that $t_2 \geq t_1$. This means that to calculate the full Green's function $G^>_{ji}$ we should sum over the four processes for which the first and the second vortex tunneling events happen at the right or the left tunneling point. The amplitudes of the two processes in which vortex tunneling events occur at different points are related by changing $x_t \rightarrow -x_t$ and $x_b \rightarrow -x_b$. Likewise the amplitudes of the processes in which both evens occur at the same tunneling point can be obtained from the amplitude of the process with vortex tunneling at different points by setting $x_t = x_b =0$ and setting $L_t\to L_t\pm x_t$ and $L_b\to L_b\pm x_b$.

\subsection{Perturbative calculation of $G^>$}
\label{sec:perturbative_calc}

In the appendices we demonstrate how one can evaluate the averages of the contributions to the integrands generated in the perturbative expansion of $G_{ji}^>$. The technically simplest way of performing the calculation is to use the commutation relation between fermions and tunneling terms [see Eq.~\eqref{eq:commute_psi_T12}]
\begin{equation}
T_{12} \psi_3 = s_{13} s_{23} \psi_3 T_{12},
\end{equation}
to transform the correlation functions into one of the two forms in Eq.~\eqref{eq:canonicalforms}. The limit of large spatial separation $\Delta L\to \infty$ can then be taken using Eq.~\eqref{eq:cluster_decomposition1}. Finally we use the functional form of the correlation function of a $\psi$ and two $\sigma$'s that is fixed by conformal invariance:\cite{Francesco_Book}
\begin{equation}
\la \sigma_1 \sigma_3 \psi_i \ra=
\frac{z_{13}^{3/8}}{\sqrt{2} z_{1i}^{1/2} z_{3i}^{1/2}}.
\end{equation}
The result of this calculation is the same as the limit $\Delta L \rightarrow \infty$ of the full six-point function that can also be calculated using bosonization and a doubling trick, see App.~\ref{app:ising}.

The first non-vanishing contribution to the fermion propagator $G^>$ comes at second order in the vortex tunneling term. It is then convenient to divide the intermediate time integrals into different regions. We will use the following labeling conventions: (a) $t_1 < t_2 <0$, (b) $t_1 < 0 < t_2 < t$, and (c) $0<t_1<t_2<t$. We now calculate the contribution to the integrand from each region separately.

Let us first consider the interval $t_1 < t_2 < 0$. By straightforward expansion, and using the exchange algebra we obtain the integrand in this region
\begin{eqnarray}
I_{(a)} &=& 
\la \psi_j \psi_i T_{34} T_{12} \ra + \la T_{12} T_{34}  \psi_j \psi_i \ra 
\nonumber \\
&-& \la  T_{34} \psi_j \psi_i T_{12} \ra  - \la  T_{12} \psi_j \psi_i T_{34}  \ra 
\nonumber \\
&=& s_{i1}s_{i2} (s_{i3}s_{i4} - s_{3j}s_{4j}) \la \psi_j  T_{34} T_{12} \psi_i \ra
\nonumber \\
&-&s_{1j}s_{2j} (s_{i3}s_{i4} - s_{3j}s_{4j}) \la \psi_j T_{12}  T_{34}  \psi_i \ra.
\end{eqnarray}
The minus signs are generated when the two tunneling terms are on different Keldysh branches, i.e. when one comes from evolving forward in time and one backwards. We can simplify this expression further by noting that because of the geometry we always have $s_{i3} = s_{i1} =1$ in this region. Thus
\begin{equation}
I_{(a)} \equiv I^> = (1+s_{j4}) ( 
\la \psi_j  T_{34} T_{12} \psi_i \ra
+
s_{j2}  \la \psi_j T_{12}  T_{34}  \psi_i \ra ).
\end{equation}

Let us now consider the interval $t_1 <0 < t_2 < t$.  We denote the contribution to the integrand in this region by $I_{(b)} $. Expanding we get
\begin{eqnarray}
I_{(b)}
&=& \la T_{12} T_{34} \psi_j \psi_i  \ra + \la \psi_j T_{34} \psi_i  T_{12} \ra  
\nonumber \\
&-& \la T_{12}  \psi_j  T_{34} \psi_i  \ra - \la T_{34}  \psi_j  \psi_i T_{12} \ra
\nonumber \\
&=& \ldots = I^> .
\end{eqnarray}
To see that we get the same expression as in region (a) we have used the fact that $s_{i1}=1$ in this region. Performing the same calculation as in regions (a) and (b) for the interval  $0 < t_1 < t_2 < t$ we find that also in this region
\begin{equation}
I_{(c)} = I^> ,
\end{equation}
and hence we can use $I^>$ throughout all regions. Using cluster decomposition (i.e. taking the limit of spatial separation) and the explicit correlation functions we get the expression for the integrand. Putting back the integrals and the strength of the tunneling term we obtain the leading term in the perturbative expansion of the Green's function
\begin{multline}
G^> = \frac{\lambda_\sigma^2}{2^{3/2}} 
\int_{-\infty}^{t} dt_1 \int_{t_1}^{t} dt_2
\frac{(1+s_{j4})} {( |z_{j2}| |z_{j4}| )^{1/2} (z_{3i} z_{1i})^{1/2}}
\\
[ (1+s_{j2}) \real (z_{31}^{3/8} z_{42}^{3/8} ) -
(1-s_{j2}) \imag (z_{31}^{3/8} z_{42}^{3/8} ) ] .
\end{multline}
Note that this expression is a short form that includes a sum of many terms, it is valid for real times only and the analytic structure of the the Green's function is not apparent. It is useful to shift the time-coordinates $t_j = t - L_b - x_b - s_j$ for $j=1,2$. The resulting expression is
\begin{multline}
\label{eq:vortex_G}
G^> = \frac{\lambda_\sigma^2}{\sqrt{2}} 
\int_0^{\infty} ds_1 \int_0^{s_1} ds_2
\frac{1}{( |z_{j2}| |z_{j4}| )^{1/2} (z_{3i} z_{1i})^{1/2}}
\\
[ (1+s_{j2}) \real (z_{31}^{3/8} z_{42}^{3/8} ) -
(1-s_{j2}) \imag (z_{31}^{3/8} z_{42}^{3/8} ) ] ,
\end{multline}
where
\begin{eqnarray}
u_{j2} & = & 2 x_b + s_1 \nonumber \\
u_{j4} & = & s_2 \nonumber \\
u_{1i} &=& \tilde{t} + x_t - x_b -s_1 \nonumber \\
u_{3i} &=& \tilde{t} -x_t - x_b -s_2 \\
u_{31} &=& s_1 - s_2 - 2 x_t \nonumber \\
u_{42} &=& s_1 - s_2 + 2 x_b \nonumber \\
\tilde{t} &=& t  - L_t - L_b . \nonumber
\end{eqnarray}
Note that the dependence on the parameters $t$, $L_t$, and $L_b$ only enters in the combination $\tilde{t}$. The analytic structure is much more transparent in this equation. For tunneling at the same point, i.e. $x_t = x_b = 0$, we always have $s_{j2}=1$ and the result simplifies to
\begin{multline}
G^>_{x_b=x_t=0} = \lambda_\sigma^2 \sqrt{2} \cos \Bigl(\frac{3 \pi}{8}\Bigr)
\int_0^{\infty} ds_1 \int_0^{s_1} ds_2
\\ \times
\frac{ |z_{31}|^{3/4}}
{( |z_{j2}| |z_{j4}|)^{1/2} (z_{3i} z_{1i})^{1/2}} .
\end{multline}
From this expression we see that $\real [ G^> ] \neq 0$ only for times such that $t \geq L_t + L_b$. Since $\real [ G^> ]$ is proportional to the retarded Green function $G^R$, this is a reflection of the causality of the theory: information has to have time to propagate through the system for $G^R$ to be non-zero.

The Green's function $G^>$ has a singular part that is given by
\begin{equation}
G^> \sim \lambda_\sigma^2 T^{-3/4}[-i\log|\xi|+\pi \Theta(\xi)],\label{eq:g_singular}
\end{equation}
with $\Theta(x)$ the Heaviside step function and
\begin{equation}
\xi = T(t-L_t-L_b-x_t-x_b)\ll 1.
\end{equation}

\subsection{Conductance}

\begin{figure}
\centering
\includegraphics[width=.9\linewidth]{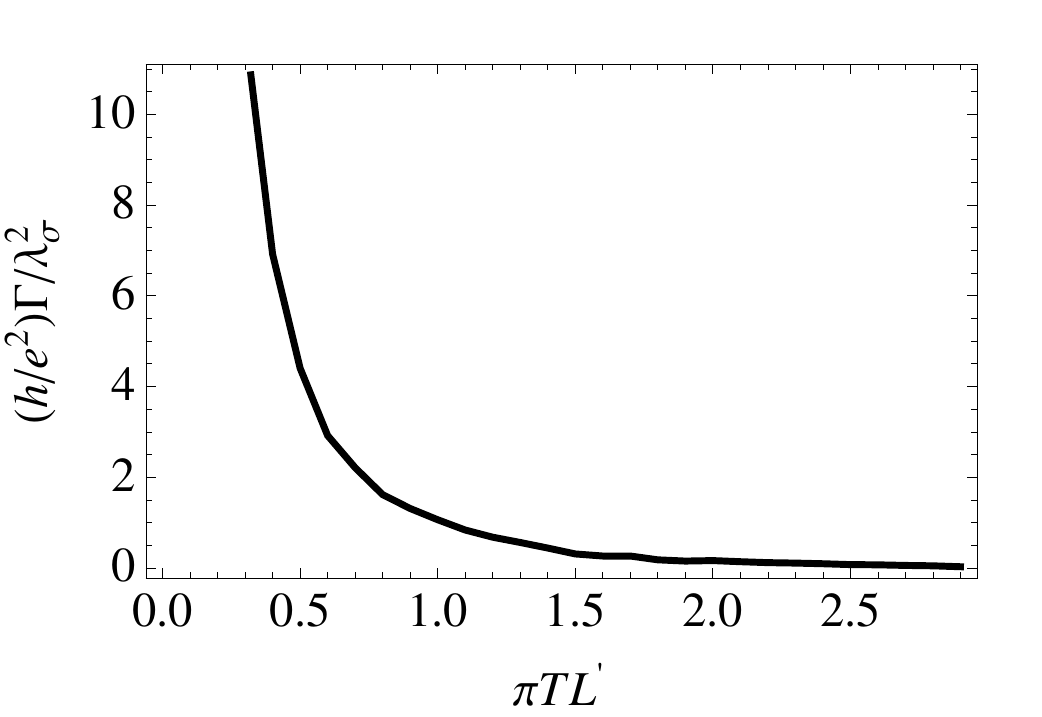}
\caption{\label{fig:conductance} Normalized conductance $(h/e^2)\Gamma/\lambda_\sigma^2\equiv T^{-7/4} F[0,0,(L_t+L_b)T,L'T]$ 
of a single quantum point contact due to vortex tunneling as a function of temperature. The parameters of the setup are $L_t=L_b=L'$.}
\end{figure}

Substituting  the propagator in Eq.~\eqref{eq:Gprime0} for the reference edge into the expression for conductance in Eq.~\eqref{eq:Gamma_v2} we obtain
\begin{multline}
\frac{\Gamma}{e^2/h} = -L' \imag [ G^> ]_{t=L'}
\\
+ \int_{0}^{\infty} dt \mathcal{P} \frac{T t}{\sinh(\pi T u_{j'i'})}
\real [ G^> ].\label{eq:conductance_vortex}
\end{multline}
Together with Eq.~\eqref{eq:vortex_G} this expression provides a closed expression determining the contribution from each process to the conductance, which may be directly evaluated numerically. Since $G^>$ only has a logarithmic divergence, the short distance cutoff $a$ may be directly set to zero in this expression.
By substituting the singular part of $G^>$ into Eq.~\eqref{eq:conductance_vortex} one can see that the conductance contribution is a continuous function of all the parameters of the problem. It may be written as
\begin{equation}
\Gamma_{LR} = \frac{e^2}{h} \frac{\lambda_\sigma^2 F[x_t T, x_b T, (L_t+L_b)T, L'T]}{T^{7/4}} ,
\end{equation}
with $F$ a universal continuous function. In the low temperature limit, when all of the arguments of $F$ are small, the contributions to conductance from vortex tunneling at different points $\Gamma_{LL}$, $\Gamma_{RR}$, $\Gamma_{LR}$, and $\Gamma_{RL}$ are all equal to each other and to
\begin{equation}
\Gamma_0 = \frac{e^2}{h} \frac{\lambda_\sigma^2 F(0, 0, 0, 0)}{T^{7/4}},\label{eq:cond_lowt}
\end{equation}
with $F(0,0,0,0)\approx 1.5$. In the other limit, when either $|x_t+x_b|T\gg 1$ or $|L_t+L_b-L'|T\gg 1$ the function $F$ is exponentially small, or in other words conductance is suppressed due to thermal averaging. We have evaluated the conductance of a single point contact due to vortex tunneling numerically with the result shown in Fig.~\ref{fig:conductance}. At low temperatures $\Gamma\times T^{7/4}\to \textrm{constant}$ as expected, and at high temperatures $\Gamma\sim \exp(-T|L'-L_t-L_b|)$.

The scaling exponent of conductance $-7/4$ is different from $-3/2$, the exponent of tunneling conductance in the 5/2 FQH effect. This naturally follows from the very different mechanisms of conduction in the two systems: current is carried by charged modes in 5/2 FQH system, while ``Dirac to Majorana converter'' forms current in topological insulators.

\section{Quasiclassical approach and fermion parity measurement}
\label{sec:topologicalcharge}

The most interesting application of the interferometer setup with vortex tunneling is that it allows for the detection of the fermion parity of the superconducting island between the two point contacts.\cite{DasSarma_topcharge2005,SternHalperin2006,Bonderson2006} This is possible because vortices acquire a phase of $\pi$ when they are moved around an odd number of fermions.\cite{Ivanov2001} In the simplest case, when there are only two bulk vortices in the central region, as shown in Fig.~\ref{fig:setup}, the interferometric signal reads out the state of the qubit formed by the bulk vortices. 

Without loss of generality we consider the case of two bulk vortices that are situated in between the left and the right tunneling regions. From the point of view of the electronic excitations the bulk vortices can be described by two localized Majorana bound states,\cite{FuKane_PRL2008} with corresponding operators $\gamma_a$ and $\gamma_b$. To describe the action of the vortex tunneling term on these excitations we include, following Ref.~\onlinecite{Rosenow_etal2008}, an extra term $\hat{P}_{ab} =i \gamma_a \gamma_b$ in the left tunneling operator. This operator captures the property that upon changing the phase of the order parameter in the superconductor by $\pm 2 \pi$ the Majorana modes localized in the vortex cores gains a minus sign.

In the absence of bulk-edge coupling the fermion parity of the vortex pair is a good quantum number that does not change with time. In that case the extra term that is added to the left tunneling term $\hat{P}_{ab}$ measures the fermion parity of the qubit defined by $\gamma_a$ and $\gamma_b$. This means that we can replace $\hat{P}_{ab} \rightarrow (-1)^{n_f}$, where $n_f$ is the number of fermions in the two vortices. In the second order calculation this factor enters only in the contributions where one vortex tunnels at the left tunneling point and one at the right, so the total conductance is equal to
\begin{equation}
\Gamma = \Gamma_{LL} + \Gamma_{RR} + (-1)^{n_f} (\Gamma_{LR}+\Gamma_{RL}).\label{eq:our_conductance}
\end{equation}
The expressions for the $\Gamma$'s were calculated in the previous section. The effect of bulk-edge coupling is presumably similar to the case of the $5/2$ FQH effect that has been studied in great detail recently.\cite{OverboschWen2007,Rosenow_etal2008,Rosenow_etal2009,BisharaNayak2009}

The phenomenological picture of the non-Abelian interferometry presented in Ref.~\onlinecite{Akhmerov_PRL_2009} can be summarized in the following way.
First an incoming electron is split into two Majorana fermions when it approaches the superconductor. Next one of these Majorana fermions is further split into two edge vortices, or $\sigma$ excitations. The edge vortices tunnel at either of the two point contacts, and recombine into a Majorana fermion again. Finally two Majorana fermions combine into electron or a hole as they leave the superconductor. At zero voltage any dynamic phases are prohibited by electron-hole symmetry, so the outgoing current may be written as
\begin{equation}
 I=\frac{e^2}{h}V \bigl[\tilde{\lambda}_{\sigma L}^{\,2} +\tilde{\lambda}_{\sigma R}^{\,2}
 +2 (-1)^{n_f}\tilde{\lambda}_{\sigma L}\tilde{\lambda}_{\sigma R}\bigr],
\label{eq:quasiclassical_conductance}
\end{equation}
where $\tilde{\lambda}_{\sigma a}$ (with $a = L,R$) is an effective vortex tunneling amplitude (here we allow for different vortex tunneling amplitudes at the left and right tunneling points).

Comparing Eqs.~\eqref{eq:cond_lowt}-\eqref{eq:our_conductance} with Eq.~\eqref{eq:quasiclassical_conductance} we see that at low temperatures the effective vortex tunneling amplitude is equal to
\begin{equation}
 \tilde{\lambda}_{\sigma a}= \lambda_{\sigma a}T^{-7/8}\sqrt{F(0,0,0,0)}.\label{eq:effective_vortex}
\end{equation}
Once this identification is done, the quasiclassical picture is directly applicable given that $1/T$ is much larger than the characteristic length of the system and the second order perturbation theory still holds ($\tilde{\lambda}_{\sigma a} \ll 1$).

\section{Conclusions}
\label{sec:conclusions}

In this paper we have introduced a theory for a non-Abelian interferometer on the surface state of a 3D topological insulator brought in proximity to an s-wave superconductor. This theory uses CFT to describe the vortex field following Ref.~\onlinecite{Fendley2007a}, and  is an extension of the earlier qualitative discussion in Ref.~\onlinecite{Akhmerov_PRL_2009}. In particular we showed that if the temperature is low and tunneling is sufficiently weak, it is possible to introduce an effective tunneling amplitude of vortices according to Eq.~\eqref{eq:effective_vortex}. This justifies the simple quasiclassical description of vortex tunneling used in Ref.~\onlinecite{Akhmerov_PRL_2009}.

Because the vortex tunneling term is a relevant operator, the perturbative treatment is only valid at high enough temperatures. This statement is reflected in the divergence of conductance $\Gamma \sim  T^{-7/4}$. The scaling exponent $-7/4$ is different from the tunneling conductance scaling exponent $-3/2$ of the 5/2 FQH setup in the linear response regime due to the different structure of current operators in the two systems.

\acknowledgments

We acknowledge useful discussion with C.~W.~J Beenakker, C.-Y. Hou, and B.~J. Overbosch.
This research was supported by the Dutch Science Foundation NWO/FOM.
J.N. thanks the Swedish research council (vetenskapsr{\aa}det) for funding in the final stage of this project.

\appendix

\section{Vortex tunneling term}
\label{app:ising}

In this appendix we show how one can calculate the amplitude for transferring a fermion between the two edges in terms of two vortex tunneling events using bosonization with the help of a doubling trick. This is an old technique that goes back to the seventies,\cite{ZuberItzykson1977} which is now textbook material.\cite{Francesco_Book,Gogolin_Book} In the appendices we use the condensed coordinate conventions introduced in Sec.~\ref{sec:coordinates}, but we'll keep the gauge choice implied by the sign of $s_{ij}$ unspecified.

\subsection{Non-chiral extension of the system}

The logic of the procedure can be motivated as follows (see also the construction in Ref.~\onlinecite{Rosenow_etal2008}). We are interested in the tunneling of a chiral Majorana fermion between two edges of a sample (cf. Fig.~\ref{fig:singe_edge}). Because of the fermion doubling feature it is convenient to enlarge the system by adding an additional counter-propagating  chiral Majorana fermion. These two copies can then be described as the continuum limit of a lattice model of local Majorana fermions (described by lattice operators $\gamma_l^\dagger = \gamma_l$) that are allowed to hop to their nearest neighbors:
\begin{equation}
H = -t \sum_{l=1}^{2N}  i \gamma_{l} \gamma_{l+1} .
\label{eq:majoranachainH}
\end{equation}
The fermion parity operator is then $\hat{P} \equiv \prod_{l=1}^{2N}e^{i \pi/4}\gamma_l$.
This system is known to map onto the (quantum) Ising chain in a transverse field at criticality (see e.g. Ref.~\onlinecite{Gogolin_Book}), which is also equivalent to the classical 2D Ising model at its critical point. In the Ising model there are spin and disorder fields that are non-local in terms of the lattice fermions. It is easy to write down explicit expressions for the spin and disorder operators in terms of a string of Majorana fermions on the lattice, for example
\begin{subequations}
\begin{eqnarray}
\sigma_{2i+1} \sigma_{2j+1} &=&  \prod_{l=2i+1}^{2j} e^{i \pi/4} \gamma_{l} ,\\
\mu_{2i} \mu_{2j} &=&  \prod_{l=2i}^{2j-1} e^{i \pi/4} \gamma_{l} , \\
\sigma_{2i+1} \mu_{2j} &=&  e^{-i \pi/4}\prod_{l=2i+1}^{2j-1} e^{i \pi/4} \gamma_{l}.
\end{eqnarray}
\end{subequations}
It is clear from these expressions that a $\sigma \mu$ term changes the fermion parity of the system whereas $\sigma \sigma$ and $\mu \mu$ do not.

Now we are not interested in the lattice theory itself but rather the low-energy theory which is obtained in the continuum limit of the lattice model. This limit is known to map onto the Ising CFT. This is a thoroughly studied system and we can hence rely on results from the large literature on this topic.

In particular, on the lattice we know that a vortex tunneling term has to be of the form $\sigma_1 \sigma_2$ or $\mu_1 \mu_2$, otherwise the fermion parity is changed. Furthermore, from the operator product expansion of the Ising CFT:\cite{Francesco_Book,Allen_Senechal2000}
\begin{subequations}
\begin{eqnarray}
\sigma_1 \sigma_2 &\sim& \frac{1}{(z_{12} \zb_{12})^{1/8}} + 
\frac{1 }{2} (z_{12} \zb_{12})^{3/8}  i \psi_2 \bar {\psi}_2 , \\
\mu_1 \mu_2 &\sim& \frac{1}{(z_{12} \zb_{12})^{1/8}} - 
\frac{1 }{2} (z_{12} \zb_{12})^{3/8}  i \psi_2 \bar {\psi}_2 ,
\end{eqnarray}
\end{subequations}
we see that a pair of $\sigma$'s (or a pair of $\mu$'s) can change the parity of right-movers. Since our tunneling term is not allowed to do this we take the tunneling term in the non-chiral system to be $\tilde{T}_{12} \propto \sigma_1 \sigma_2 + \mu_1 \mu_2$. Clearly the parity-changing term is canceled with this choice. Another way of putting this is to say that this combination enforces the tunneling term to be in the identity channel.

It is known that two {\it independent} copies of the Ising model can be bosonized using Abelian bosonization.\cite{ZuberItzykson1977,Francesco_Book} It is then a straightforward calculation (using for example the explicit expressions in the appendix of Ref.~\onlinecite{Allen_Senechal2000}) to show that the doubled tunneling term can be bosonized as
\begin{equation}
\widetilde{T} _{12} \widetilde{T}_{12}' = \cos \Bigl(\frac{\phi_1-\phi_2}{2}\Bigr)
\cos\Bigl(\frac{\bar{\phi}_1-\bar{\phi}_2}{2}\Bigr).
\end{equation}
It is important to note that the primed system is an \emph{independent} copy of the system in this expression, and that it is introduced as a trick to allow for a simple calculation of various correlation functions.

\subsection{From non-chiral back to chiral}

Since we are only interested in the right-moving part of the tunneling term we would like to get rid of the left-moving part in the last equation. Because of the factorization of the right- and left-moving parts we are allowed to use 
\begin{equation}
T _{12} T_{12}' = \cos \Bigl(\frac{\phi_1-\phi_2}{2}\Bigr),
\end{equation}
as the doubled tunneling term in the chiral system. Here the cosines are to be understood as shorthands for $\cos(a-b) = (e^{ia}e^{-ib} + e^{-ia}e^{ib})/2$. The exponentials in these expressions are actually dimensionful vertex operators, see e.g. Ref.~\onlinecite{Boson_beginners} for a detailed discussion. With this representation together with the bosonized representation of the Majorana fermion in the unprimed system
\begin{equation}
\psi_i = \sqrt{2} \cos(\phi_i) ,
\end{equation}
and the standard bosonization formula (which holds if $\sum_{i=1}^N \alpha_i = 0$, otherwise the expectation value vanishes)
\begin{equation}
\la e^{i \alpha_1 \phi_1} e^{i \alpha_2 \phi_2} \cdots e^{i \alpha_N \phi_N}\ra
= \prod_{1\leq i<j \leq N} z_{ij}^{\alpha_i \alpha_j},
\end{equation}
with
\begin{equation}
z_{ij} = \frac{\sin[\pi T(a+i u_{ij})]}{\pi T},
\label{eq:zdef}
\end{equation}
we can in principle calculate any correlation function using the bosonization formalism. In particular we can calculate the full six-point function including two $\psi$'s and two tunneling terms. This will be done in the next subsection, but let us first check that the representation reproduces known results for the 2-, 3- and 4-point functions.

Let us first consider the vortex 2-point function. This is calculated via
\begin{equation}
\la T _{12}\ra^2 = \la T _{12} T_{12}' \ra
= \frac{1}{z_{12}^{1/4}}.
\end{equation}
Taking the square root we obtain the correct result for a field with dimension $\frac{1}{16}$
\begin{equation}
\label{eq:averageT12}
\la T _{12}\ra = \frac{1}{z_{12}^{1/8}}.
\end{equation}
Similarly the fermion two-point function is $\la \psi_i \psi_j \ra = z_{ij}^{-1}$. The vortex 4-point function can be computed from
\begin{equation}
\la T_{12} T_{34} \ra^2 = \frac{1}{2}
\Bigl[ \Bigl( \frac{z_{13} z_{24} }{z_{12} z_{23} z_{34} z_{14}} \Bigr)^{1/4}
+ \Bigl( \frac{z_{14} z_{23} }{z_{13} z_{24} z_{12} z_{34}} \Bigr)^{1/4} \Bigr]
.
\label{eq:spin4point}
\end{equation}
Taking the square root of this expression we get the known correlation function of four $\sigma$'s for which $\sigma_1$ and $\sigma_2$ fuse to the identity.\cite{BPZ1984,Ginsparg88} Now we use the conventions from the main part of the paper and take the limit $\Delta L \rightarrow \infty$. In this case only one of the terms in Eq.~\eqref{eq:spin4point} survives and
\begin{equation}
\label{eq:T2ave1}
\la T_{12} T_{34} \ra \underset{\Delta L \rightarrow \infty}{=} \frac{1}{\sqrt{2}}
\Bigl( \frac{z_{14} z_{23} }{z_{13} z_{24} z_{12} z_{34}} \Bigr)^{1/8} .
\end{equation}
We also have
\begin{equation}
\la \psi_i T_{12} \ra =0 ,
\end{equation}
which is consistent with the notion that the tunneling of a vortex cannot create a fermion (or equivalently change the fermion parity). It is also straightforward to show that
\begin{equation}
\la \psi_i \psi_j T_{12} \ra  \underset{\Delta L \rightarrow \infty}{=} 0 ,
\end{equation}
which means that a single vortex tunneling event is not enough to be able to transfer a fermion between the two edges.

\subsection{The six-point function}

To calculate the contribution from a tunneling of two vortices we need the six-point function of two $\psi$'s and four $\sigma$'s. This correlation function is a special case of the more general one that was first calculated in Ref.~\onlinecite{NayakWilczek} with a similar method. To calculate the six-point function we use
\begin{multline}
\la \psi_i \psi_j T_{12} T_{34} \ra \la T_{12}'  T_{34}' \ra
\\
= 2 \Bigl< \cos(\phi_i) \cos(\phi_j) 
\cos \Bigl(\frac{\phi_1-\phi_2}{2}\Bigr) \cos \Bigl(\frac{\phi_3-\phi_4}{2}\Bigr)
\Bigr>.
\\
= \frac{1}{4 z_{ij} (z_{12} z_{34})^{1/4}}
\\ \times
\Bigl\{
\Bigl[ \Bigl( \frac{z_{i1} z_{i3} z_{j2} z_{j4}}{z_{i2} z_{i4} z_{j1} z_{j3}} \Bigr)^{1/2}
+ (i \leftrightarrow j) \Bigr] \Bigl( \frac{z_{13} z_{24} }{z_{14} z_{23} } \Bigr)^{1/4}
\\
+
\Bigl[ \Bigl( \frac{z_{i1} z_{i4} z_{j2} z_{j3}}{z_{i2} z_{i3} z_{j1} z_{j4}} \Bigr)^{1/2}
+ (i \leftrightarrow j) \Bigr] \Bigl( \frac{z_{14} z_{23} }{z_{13} z_{24} } \Bigr)^{1/4}
\Bigr\}.
\end{multline}
Dividing this with the square root of Eq.~\eqref{eq:spin4point} the result agrees with that of Ref.~\onlinecite{NayakWilczek}. We now take the limit of spatial separation $\Delta L \rightarrow \infty$, the only one term that remains is
\begin{multline}
\la \psi_i \psi_j T_{12} T_{34} \ra \la T_{12}'  T_{34}' \ra
\\
\underset{\Delta L \rightarrow \infty}{=}
\frac{(z_{13}z_{24})^{1/4}}{4 (z_{i1} z_{i3} z_{j2} z_{j4})^{1/2}}
\times
\frac{(z_{i2} z_{i4} z_{j1} z_{j3})^{1/2}}{z_{ij} (z_{12} z_{34} z_{14} z_{23})^{1/4}}.
\end{multline}
Combining this with Eq.~\eqref{eq:T2ave1} we find
\begin{equation}
\label{eq:averageresult1}
\la \psi_i \psi_j T_{12} T_{34} \ra = 
\frac{(z_{13}z_{24})^{3/8}}{2^{3/2} (z_{i1} z_{i3} z_{j2} z_{j4})^{1/2} }
\times
\frac{(z_{i2} z_{i4} z_{j1} z_{j3})^{1/2}}{z_{ij} (z_{14} z_{23})^{1/2}},
\end{equation}
To get this result we have removed the phases associated with $z_{12}^{-1/8}$ and  $z_{34}^{-1/8}$. These phases are canceled when one makes sure that the tunneling term is described by a Hermitean term in the Hamiltonian. This is exactly the phase of $\la T_{12}\ra$ in Eq.~\eqref{eq:averageT12}.

Other orderings of the fermions and the tunneling terms are obtained by exchanging the indexes, for example
\begin{equation}
\la \psi_i  T_{12} \psi_j T_{34} \ra = 
\frac{(z_{13}z_{24})^{3/8}}{2^{3/2} (z_{i1} z_{i3} z_{2j} z_{j4})^{1/2} }
\times
\frac{(z_{i2} z_{i4} z_{1j} z_{j3})^{1/2}}{z_{ij} (z_{14} z_{23})^{1/2}}.
\end{equation}
The indexes on the $z$'s should have the same order as they appear in in the original expression. This prescription was used in e.g. Ref.~\onlinecite{OverboschWen2007} and is equivalent to the Keldysh formalism for chiral bosons which is reviewed in e.g. Refs.~\onlinecite{Chamon1996a} and \onlinecite{Kim2006a}. In the limit of spatial separation the last term gives a phase factor that depends on the order of the tunneling terms and the fermions according to
\begin{eqnarray}
\label{eq:alphadef}
e^{i \alpha_p} = -i s_{ij}
\begin{cases}
1, & p = ij1234 , \; 1234ij , \; 12ij34 \\
-1, & p = i1234j \\
i s_{ij}, & p = i12j34 , \; 12i34j .
\end{cases}
\end{eqnarray}

\section{Exchange algebra}
\label{app:exchange}

An alternative formalism is provided by the exchange algebra of Ref.~\onlinecite{RehrenSchroer}. In this formalism the action of the spin field is described by two types of operators $a$ and $b$ and their conjugates. $a$ creates an excitation with dimension $\frac{1}{16}$ when acting on the vacuum, which is denoted by the shorthand $a |0\ra = |\frac{1}{16}\ra$. The conjugate $a^{\dagger}$ interpolates in the opposite direction: 
$a^\dagger|\frac{1}{16}\ra = |0\ra$. Similarly $b$ and $b^\dagger$ interpolates between states of dimensions $\frac{1}{16}$ and $\frac{1}{2}$ according to $b |\frac{1}{16}\ra = |\frac{1}{2}\ra$ and $b^\dagger |\frac{1}{2}\ra = |\frac{1}{16}\ra$. The exchange algebra is described by the following relations
\begin{equation}
\label{eq:algebraRS1}
\begin{pmatrix}
a_1 a_2^\dagger \\ b_1^\dagger b_2
\end{pmatrix}
= \frac{e^{i s_{12} \frac{\pi}{8}}}{\sqrt{2}}
\begin{pmatrix}
1 & e^{- i s_{12} \frac{\pi}{2}} \\ e^{- i s_{12} \frac{\pi}{2}} & 1
\end{pmatrix}
\begin{pmatrix}
a_2 a_1^\dagger \\ b_2^\dagger b_1
\end{pmatrix},
\end{equation}
\begin{subequations}
\label{eq:algebraRS2}
\begin{eqnarray}
a_1^\dagger a_2 &=& e^{-i s_{12} \frac{\pi}{8} } a_2^\dagger a_1 ,\\
b_1 b_2^\dagger &=& e^{-i s_{12} \frac{\pi}{8} } b_2 b_1^\dagger   ,\\
b_1 a_2 &=& e^{-i s_{12} \frac{\pi}{8} } e^{i s_{12} \frac{\pi}{2} } b_2 a_1, \\
a_1^\dagger b_2^\dagger &=& e^{-i s_{12} \frac{\pi}{8} } e^{i s_{12} \frac{\pi}{2} }  a_2^\dagger b_1^\dagger .
\end{eqnarray}
\end{subequations}
The tunneling operator, e.g. $T_{12}$, consists of a product of two $\sigma$'s in the identity channel, which we denote $[\sigma_1 \sigma_2]_I$. When acting on states with dimension $0$ or $\frac{1}{2}$ this implies that we are allowed to use the representations
\begin{equation}
\label{eq:identitychannelRS1}
[\sigma_1 \sigma_2]_I \rightarrow
\begin{cases}
a_1^\dagger a_2 , \quad |0\ra \rightarrow |0\ra \\
b_1 b_2^\dagger , \quad |\frac{1}{2}\ra  \rightarrow |\frac{1}{2}\ra  
\end{cases}.
\end{equation}
Another important point is that the tunneling term should be represented by a Hermitean term in the Hamiltonian. This can be achieved by explicitly adding the Hermitean conjugate in the definition of the tunneling term:
\begin{equation}
T_{12} \propto [\sigma_1 \sigma_2 +\sigma_2 \sigma_1 ]_I =
\begin{cases}
(1+ e^{-i s_{12} \frac{\pi}{8}}) [\sigma_2 \sigma_1]_I \\
(1+ e^{i s_{12} \frac{\pi}{8}}) [\sigma_1 \sigma_2]_I 
\end{cases}.
\end{equation}
In the last step we used Eqs.~\eqref{eq:identitychannelRS1} and \eqref{eq:algebraRS2}. By adjusting the amplitude to conform with the result of the previous section [see discussion below Eq.~\eqref{eq:averageresult1}] we define
\begin{equation}
T_{12} = e^{-i s_{12} \frac{\pi}{16}} [\sigma_2 \sigma_1]_I = e^{i s_{12} \frac{\pi}{16}} [\sigma_1 \sigma_2]_I ,
\end{equation}
which is Hermitean.

Similarly we can represent the fermion field in terms of $a$'s and $b$'s with coinciding arguments\cite{psiabcomment}
\begin{equation}
\label{eq:psiRS1}
\psi_1 \propto
\begin{cases}
 b_1 a_1 , \quad |0\ra \rightarrow |\frac{1}{2}\ra \\
 a_1^\dagger b_1^\dagger , \quad |\frac{1}{2}\ra \rightarrow |0\ra
\end{cases} .
\end{equation}
Using Eqs.~\eqref{eq:psiRS1} and \eqref{eq:identitychannelRS1} together with the exchange algebra of Eqs.~\eqref{eq:algebraRS1} and \eqref{eq:algebraRS2} it is straightforward to show that in all cases we have the following commutation relations
\begin{equation}
[\sigma_1\sigma_2]_I \psi_3 = s_{13} s_{23} \psi_3 [\sigma_1 \sigma_2 ]_I ,
\end{equation}
which immediately implies the commutation relation between a tunneling term and a fermion is
\begin{equation}
\label{eq:commute_psi_T12}
T_{12} \psi_3 = s_{13} s_{23} \psi_3 T_{12}.
\end{equation}
With this very important relation we can always transform the correlation functions that we want to calculate (see Sec.~\ref{sec:perturbative_calc}) into one of two different forms:
\begin{subequations}
\label{eq:canonicalforms}
\begin{eqnarray}
\la \psi_j  T_{12} T_{34} \psi_i \ra , \\
\la \psi_j  T_{34} T_{12} \psi_i \ra .
\end{eqnarray}
\end{subequations}
Using the exchange algebra we can cluster decompose the last two expressions, in the limit of spatial separation we are left with
\begin{subequations}
\label{eq:cluster_decomposition1}
\begin{eqnarray}
\la \psi_j  T_{12} T_{34} \psi_i \ra
\underset{\Delta L \rightarrow \infty}{=}  \frac{e^{-i s_{12} \frac{\pi}{2} }}{\sqrt{2}}
\la \psi_j b_2 a_4 \ra \la a_1^\dagger b_3^\dagger \psi_i \ra , \\
\la \psi_j  T_{34} T_{12} \psi_i \ra
\underset{\Delta L \rightarrow \infty}{=} \frac{e^{-i s_{12} \frac{\pi}{2} } }{\sqrt{2}}
\la \psi_j b_4 a_2 \ra \la a_3^\dagger b_1^\dagger \psi_i \ra .
\end{eqnarray}
\end{subequations}

We have checked that the result of the formalism in this appendix gives identical results to those of the formalism in App.~\ref{app:ising}. Although the exchange algebra is derived at $T=0$ it also holds at finite temperatures.

\bibliography{topological}

\begin{thebibliography}{39}
\bibitem[{gau({\natexlab{b}})}]{gauge_choice_comment2}
\bibinfo{note}{Other gauges choices are possible, but are more cumbersome to
  use in the calculation.}
\bibitem[{psi()}]{psiabcomment}
\bibinfo{note}{The correct expression is obtained in the properly scaled limit
  of coinciding coordinates. For example, to go from $|\frac{1}{2}\ra
  \rightarrow |0\ra$ we may use $\psi(u) = \lim_{\delta \rightarrow 0^+} C
  \delta^{-3/8} a^\dagger(u+\delta) b^\dagger(u)$. Here $C$ is a constant that
  can be determined by fixing the normalization, but this is not necessary to
  derive the commutation relations.}
\end{thebibliography}

\end{document}